\documentclass[12pt]{article}
\usepackage{amsmath}
\usepackage{cite}
\usepackage{epsfig}
\usepackage{color}
\usepackage{amssymb}

\newdimen\nude\newbox\chek
\def\slash#1{\setbox\chek=\hbox{$#1$}\nude=\wd\chek#1{\kern-\nude/}}

\begin{document}

\def\la{\mathrel{\mathpalette\fun <}}
\def\ga{\mathrel{\mathpalette\fun >}}
\def\fun#1#2{\lower3.6pt\vbox{\baselineskip0pt\lineskip.9pt
  \ialign{$\mathsurround=0pt#1\hfil##\hfil$\crcr#2\crcr\sim\crcr}}}

\def\abs#1{\left|#1\right|}
\def\bc{\begin{center}}
\def\ec{\end{center}}
\def\emi{\end{minipage}}
\def\epi{\end{picture}}
\def\bearray{\begin{eqnarray}}
\def\eearray{\end{eqnarray}}
\def\cO#1{{{\cal{O}}}\left(#1\right)}
\def\mL{\langle \,\ell\, \rangle}
\def\eps{\epsilon}
\newcommand{\im}{\mathrm{Im\,}}
\newcommand{\cl}{\underline{\lambda}}
\newcommand{\mh}{m_h}
\newcommand{\be}{\begin{equation}}
\newcommand{\ee}{\end{equation}}
\newcommand\ccb{{c\bar{c}}}
\newcommand\xf{$x_F$ }
\newcommand\jps{{J/\psi}}
\newcommand\psp{\psi'}
\newcommand\ps{\psi }
\newcommand\E{\epsilon}
\newcommand\Dbar{\bar{D}}
\newcommand\Et{{\tilde{E}}}
\newcommand\epst{{\tilde{\epsilon}}}

 \newskip\humongous \humongous=0pt plus 1000pt minus 1000pt
 \def\caja{\mathsurround=0pt} \def\eqalign#1{\,\vcenter{\openup1\jot
 \caja   \ialign{\strut \hfil$\displaystyle{##}$&$
 \displaystyle{{}##}$\hfil\crcr#1\crcr}}\,} \newif\ifdtup
 \def\panorama{\global\dtuptrue \openup1\jot \caja
 \everycr{\noalign{\ifdtup \global\dtupfalse     \vskip-\lineskiplimit
 \vskip\normallineskiplimit      \else \penalty\interdisplaylinepenalty \fi}}}
 \def\eqalignno#1{\panorama \tabskip=\humongous
 \halignto\displaywidth{\hfil$\displaystyle{##}$
 \tabskip=0pt&$\displaystyle{{}##}$\hfil
 \tabskip=\humongous&\llap{$##$}\tabskp=0pt     \crcr#1\crcr}}

\newcounter{hran}
\renewcommand{\thehran}{\arabic{hran}}

\def\bmini{\setcounter{hran}{\value{equation}}
\refstepcounter{hran}\setcounter{equation}{0}
\renewcommand{\theequation}{\thehran\alph{equation}}\begin{eqnarray}}

\def\bminiG#1{\setcounter{hran}{\value{equation}}
\refstepcounter{hran}\setcounter{equation}{-1}
\renewcommand{\theequation}{\thehran\alph{equation}}
\refstepcounter{equation}\label{#1}\begin{eqnarray}}

%
%
\def\emini{\end{eqnarray}\relax\setcounter{equation}{\value{hran}}\renewcommand{\theequation}{\arabic{equation}}}

\begin{flushright}
ECT*-03-13\\
hep-ph/0306235\\
June 2003
\end{flushright}
\vskip1.5cm

\begin{center}
{\Large\bf Quenching of hadron spectra in DIS \\[0.4cm] on nuclear targets}
\end{center}

\begin{center}
{\large  Fran\c{c}ois Arleo}\\[0.5cm]
{\em ECT* and INFN, G.C. di Trento,} \\
{\em Strada delle Tabarelle, 286} \\
{\em 38050 Villazzano (Trento), Italy}
\end{center}

\begin{abstract}
The multiple scatterings incurred by a hard quark produced in a nuclear medium induce the emission of soft gluons which carry a fraction of the quark energy and eventually affect the hadronization process. Here, the depletion of semi-inclusive hadron spectra in DIS on various nuclei (N, Ne, Cu, Kr) is computed as a function of $\nu$ and $z$ to leading order in $\alpha_s$ through medium-modified fragmentation functions. Using the transport coefficient $\hat{q}$ previously determined from Drell-Yan production, the predictions are found to be in good agreement with EMC and HERMES preliminary data. Calculations on Xe targets are also presented and discussed.
\end{abstract}

\section{Introduction}

The interest of semi-inclusive hadron production in Deep Inelastic Scattering (DIS) on nuclear targets is at least twofold. First, and on rather general grounds, this reaction should help us to understand how a given QCD medium affects the non-perturbative hadronization mechanism. In particular, the use of nuclei ---~with a size $R$ of a few fm~--- provides a direct information on the soft time scales $\Lambda \sim 1/R$ involved in the dynamics of hadronization. In the second place, such a process also allows one to gain a quantitative insight into the scattering properties of cold nuclear matter. These may serve as a baseline to which properties of hot QCD media should be contrasted. This is clearly an important issue in high energy heavy ion physics where the quark gluon plasma is expected to be formed.

Measurements of semi-inclusive hadron electroproduction on nuclei were performed a decade ago by the European Muon Collaboration (EMC) experiment who reported a significant attenuation of charged hadron spectra in Copper as compared to Deuterium targets~\cite{Ashman:1991cx}. This was a strong indication that quark fragmentation functions get modified in a nuclear environment\footnote{The attenuation of hadron electroproduction on a Carbon target was actually first observed at SLAC~\cite{Osborne:1978ai}. However, the SLAC measurements were {\it not} normalized to the number of inclusive events $N_A^e$ (see Eq.~(\ref{eq:multDIS})) but the atomic mass number, hence rather reflected the nuclear modifications of parton densities (see discussion in Section~\ref{se:xs}). For this reason, these data are not considered in the present study.}. More recently, the HERMES collaboration at DESY reported on high statistics measurements of charged hadron ($h^\pm$), pion ($\pi^\pm$), kaon ($K^\pm$), proton and antiproton ($p$, $\bar{p}$) electroproduction on several nuclei (Deuterium, Nitrogen, Neon, Krypton) and on a wide kinematic acceptance~\cite{Airapetian:2002ksMuccifora:2001znMuccifora:2002eaDiNezza,Elbakyan}. These two data sets make quantitative analyses possible. 

Several suggestions have been advanced so far to account for the experimental results. Models based on the nuclear absorption of the produced hadron~\cite{Bialas:1983kn,Bialas:1987cf,Accardi:2002tv,Falter:2003di}, gluon bremsstrahlung~\cite{Kopeliovich:1995jt}, or partial deconfinement in nuclei~\cite{Jaffe:1984zwClose:1985znNachtmann:1984py,Accardi:2002tv}, were shown to describe the trend of the data fairly well. Following Wang and Wang~\cite{Wang:2002ri}, we shall assume here that the observed hadron quenching actually comes from the energy loss incurred by hard quarks traveling through the nuclear medium. Although the effects of quark energy loss in DIS on nuclear targets have already been investigated in Refs.~\cite{Wang:2002ri,Arleo:2002kh,Arleo:2002ki}, a more complete analysis of EMC and HERMES data is still lacking. This is the aim of the present study.

The outline of the paper is as follows. Hadron electroproduction in nuclei is computed to leading order in the coupling constant (Section~\ref{se:xs}) using nuclear fragmentation functions modeled in Section~\ref{se:nff}. Predictions on hadron attenuation are then systematically compared to EMC and HERMES preliminary data as a function of $\nu$ and $z$ (Section~\ref{se:results}). Calculations in Xenon targets in the HERMES kinematic acceptance will also be presented. Finally, we shall discuss and summarize our results in the last Section.

\section{Hadron production in DIS on nuclei}\label{se:xs}

Hadron production in semi-inclusive DIS can be computed in perturbation theory within the QCD improved parton model. To leading order (LO) in the strong coupling constant $\alpha_s$, the multiplicity $N_A^h$ of a specific hadron species $h$ in lepton-nucleus reactions is expressed in terms of nuclear parton densities $f_q^A$ ($f_{\bar{q}}^A$) and the $q\to h$ ($\bar{q}\to h$) fragmentation functions, $D_q^h$ ($D_{\bar{q}}^h$). Normalizing it to the number of DIS inclusive events $N_A^e$ , it is given by
\begin{eqnarray}\label{eq:multDIS}
\frac{1}{N_A^e}\frac{dN_A^h(\nu,z)}{d\nu\,dz} & = & \int \, dx\, \sum_{q,\,\bar{q}} \, e^2_q\, \left[ Z\,f_q^{p/A}(x, Q^2) + (A-Z) \,f_q^{n/A}(x, Q^2) \right] \, \sigma^{\gamma^* q}(x,\nu) \, D_q^h(z, Q^2, A) \nonumber \\
& \biggm/ & \int \, dx\, \sum_{q,\,\bar{q}} \, e^2_q\, \left[ Z\,f_q^{p/A}(x, Q^2) + (A-Z) \,f_q^{n/A}(x, Q^2) \right] \, \sigma^{\gamma^* q}(x,\nu)
\end{eqnarray}
where $\nu$ and $Q$ are respectively the energy and the virtuality of the hard photon, and $z = E_h / \nu$ is the energy fraction of the virtual photon carried away by the leading hadron. The nucleus is characterized by its atomic mass number $A$ and its number of protons $Z$. The LO $\gamma^* q$ cross section appearing in (\ref{eq:multDIS}) is given by
\begin{equation}
\sigma^{\gamma^* q}(x,\nu) = \frac{4\pi \alpha_{em} \,M \,x}{Q^4} \times \left[ 1+\left(1-\frac{Q^2}{x\,s}\right)^2 \right],
\end{equation}
where $\alpha_{em}$ is the fine structure constant, $M$ the nucleon mass, and $s$ the center-of-mass energy squared of the lepton-nucleon reaction. The integral over Bjorken $x=Q^2/2M\nu$ in Eq.~(\ref{eq:multDIS}) is given by the $Q^2$ acceptance of the experiment.

At small $x \lesssim 0.05$ and low $Q^2$, the parton distribution functions (PDF) in nuclei $f_q^A$ differ significantly to those measured in a free proton, $f_q^p$~\cite{Arneodo:1994wfGeesaman:1995ydPiller:1999wx}. These so-called shadowing corrections may therefore affect the hadron production cross section~(\ref{eq:multDIS}). In the following, however, we shall not consider such corrections and assume $f_q^{p/A} = f_q^p$ for two reasons. First, the data we would like to discuss cover a Bjorken $x$ range, $x \simeq 0.1$, where the effects of nuclear shadowing prove rather small. Furthermore, the {\it normalized} hadron multiplicity Eq.~(\ref{eq:multDIS}) is rather insensitive to the absolute magnitude of the parton densities. As noted in~\cite{Accardi:2002tv}, the shadowing corrections mostly cancel out in the ratio~(\ref{eq:multDIS}) and finally affect hadron production by a percent at most.

The parton densities in a proton $f_q^p$ we use in the present analysis are given by the MRST 2001 LO parameterization~\cite{Martin:2002dr}. It has been checked however that similar results were obtained using either the GRV LO~\cite{Gluck:1998xa} and CTEQ5L~\cite{Lai:1999wy} parton distribution functions. The neutron PDF $f_q^n$ are directly deduced from $f_q^p$ by isospin conjugation, i.e.  $u^p = d^n$, $d^p = u^n$, $\bar{u}^p = \bar{d}^n$, $\bar{d}^p = \bar{u}^n$, and $\bar{s}^p = \bar{s}^n$.

In this paper, we would like to discuss the nuclear dependence of semi-inclusive hadron production. In particular, the theoretical calculations will be compared to the hadron production {\it ratio} in a heavy nucleus $A$ with respect to a (light) Deuterium target,
\begin{equation}\label{eq:suppDIS}
R_A^{h}(z,\nu) = \frac{1}{N_A^e}\,\frac{dN_A^h(z,\nu)}{d\nu\,dz}\Biggm/\frac{1}{N_D^e}\,\frac{dN_D^h(z,\nu)}{d\nu\,dz},
\end{equation}
which has been reported experimentally. When $x$ is not too small, hadron production (\ref{eq:multDIS}) is dominated by the fragmentation of valence quarks. Assuming for simplicity that only the up quark channel contributes to the process, the double production ratio Eq.~(\ref{eq:suppDIS}) approximately reduces to
\begin{equation}\label{eq:suppDIS_approx}
R_A^{h}(z,\nu) \simeq D_u^h(z, Q^2, A) \biggm/ D_u^h(z, Q^2, D).
\end{equation}
Therefore, hadron production in DIS on nuclei is mostly sensitive to the nuclear dependence of the fragmentation functions. In particular, the reported depletion of hadron multiplicity on heavy nuclei $R_A^{h}(z,\nu) < 1$ on a wide kinematic range is a clear hint that fragmentation functions are somewhat modified in the medium. 

The starting point to determine the hadron attenuation $R_A^{h}(z,\nu)$ is therefore the computation of nuclear fragmentation functions $D(z, Q^2, A)$. This task is carried out in the following section.

\section{Nuclear fragmentation functions}\label{se:nff}

\subsection{Model for medium-modified fragmentation functions}

To leading order in the coupling and at leading twist, the virtual photon picks up a quark\footnote{In the following discussion, we shall ignore the fragmentation of antiquarks as $x$ is not too small. The theoretical computations Eq.~(\ref{eq:multDIS}) consider of course this channel as well.} in the nucleus which subsequently hadronizes into the observed leading hadron. In presence of a QCD medium, however, the hard quark suffers multiple scattering and radiates soft gluons all along its path. Due to this medium-induced gluon radiation, the quark energy is reduced from $E=\nu$ to $E=\nu-\eps$ at the time of hadronization. This picture is schematically depicted in Figure~\ref{fig:model}.

\begin{figure}[h]
\begin{center}
\includegraphics[width=9.8cm]{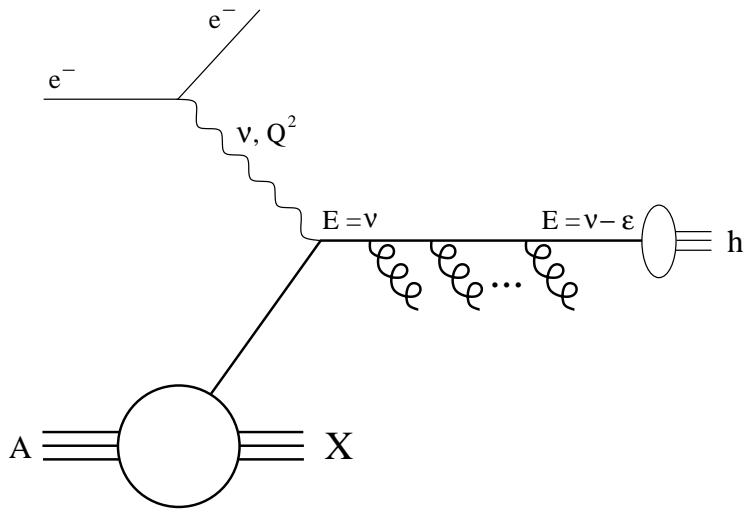}
\caption{Schematic representation of the semi-inclusive hadron electroproduction in a nuclear target. The virtual photon couples to a hard quark which subsequently emits soft gluons while escaping the nucleus.}
\label{fig:model}
\end{center}
\end{figure}

How to relate these final state interactions to nuclear fragmentation functions remains however unclear. Such an attempt has been performed recently in a series of papers~\cite{Wang:2002ri,Guo:2000nzWang:2001ifOsborne:2002dx} in which a higher-twist perturbative framework has been applied successfully to describe hadron production in DIS and heavy ion reactions. 

Here, we shall rather adopt the effective model suggested in Ref.~\cite{Wang:1996yh}. Within this approach, the quark energy shift leads to a rescaling of the momentum fraction in presence of a QCD medium,
\begin{equation}\label{eq:shift}
z = \frac{E_h}{\nu} \qquad \to \qquad z^* = \frac{E_h}{\nu - \eps} = \frac{z}{1 - \eps/\nu},
\end{equation}
where $E_h$ is the measured hadron energy and $\eps$ the energy lost by the hard quark while going through the nucleus. Consequently, the fragmentation functions in nuclei $D_q^h(z, Q^2, A)$ may simply be expressed as a function of the standard (vacuum) fragmentation functions $D_q^h(z, Q^2)$ through~\cite{Wang:1996yh}
\begin{equation}\label{eq:modelFF}
z\,D_q^h(z, Q^2, A) = \int_0^{\nu - E_h} \, d\eps \,\,D(\eps, \nu)\,\,\, z^*\,D_q^h(z^*, Q^2).
\end{equation}
Here, $D(\eps, \nu)$ denotes the probability for a quark with energy $E=\nu$ to lose an energy $\eps$. This probability distribution, or quenching weight, will be made explicit hereafter. Since fragmentation functions fall steeply at large $z$, even the small shift $\Delta z=z^*-z\approx z\,\eps/\nu$ in (\ref{eq:shift}) may substantially suppress the hadron yields in nuclear targets.

Note that in the Bjorken limit $\nu \gg \eps$, hence $Q \gg \eps$ at finite $x$, the effects of the final state interactions become negligible, $z\simeq~z^*$. No more nuclear dependence of the fragmentation functions is observed $D_q^h(z, Q^2, A)=D_q^h(z, Q^2)$, in agreement with QCD factorization theorems~\cite{Brock:1995sz}.

The vacuum fragmentation functions $D_q^h(z, Q^2)$ into various hadron species (charged hadrons, pions, kaons) which enter Eq.~(\ref{eq:modelFF}) are here given by the leading order analysis of $e^+ e^-$ data performed by Kretzer~\cite{Kretzer:2000yf}.

Let us now turn to the computation of the quenching weights for hard quarks going through nuclear matter.

\subsection{Quenching weight}

As apparent in Eq.~(\ref{eq:modelFF}), the knowledge of the probability distribution $D(\eps,\nu)$ is essential to determine the nuclear fragmentation functions. This point has been raised recently by Baier, Dokshitzer, Mueller, and Schiff (BDMS) in Ref.~\cite{Baier:2001yt} in which they express $D(\eps,\nu)$ as a Poisson series
\begin{equation}  \label{eq:poisson}
D(\eps,\nu) = \sum^\infty_{n=0} \, \frac{1}{n!} \,
\left[ \prod^n_{i=1} \, \int \, d\omega_i \, \frac{dI(\omega_i,\nu)}{d\omega} 
\right] \delta \left(\eps - \sum_{i=1}^n  \omega_i\right)
\, \,\exp \left[ - \int_0^{+\infty} d\omega \, \frac{dI(\omega, \nu)}{d\omega} \right], 
\end{equation}
assuming gluon emissions to be independent.

The expression Eq.~(\ref{eq:poisson}) explicitly relates the probability distribution in the energy loss to the gluon spectrum $dI/d\omega$ radiated by hard quarks produced in QCD media, which has been computed perturbatively by Baier, Dokshitzer, Mueller, Peign\'e, and Schiff (BDMPS). To first order in the quark energy $\cO{1/\nu}$, it reads~\cite{Baier:1998kqBaier:1997kr}
\begin{equation}\label{eq:dIdo_out}
\frac{dI(\omega,\nu)}{d\omega} = \frac{\alpha_s \, C_F}{\pi\,\omega} \left(1-\frac{\omega}{\nu}\right) \ln \left[\,\cosh^2 \sqrt{\frac{\omega_c}{2\,\omega}} - \sin^2 \sqrt{\frac{\omega_c}{2\,\omega}} \,\right] \Theta(\nu-\omega) , 
\end{equation}
where $C_F = 4/3$ is the Casimir operator in the fundamental representation and $\alpha_s = g^2 / 4\pi \simeq 1/2$ the strong coupling constant. Note that $\alpha_s$ in~(\ref{eq:dIdo_out}) is evaluated at a soft scale given by the virtuality of the radiated gluons, while the renormalization scale appearing in the parton densities and fragmentations functions in~(\ref{eq:multDIS}) is given by the photon virtuality, $Q$.  The appearance of these two scales follows our implicit assumption on the factorization between the production process and the final state interactions. The typical gluon energy in the spectrum is set by the relevant scale
\begin{equation}
\omega_c = \frac{1}{2}\,\hat{q}\, L^2
\end{equation}
which characterizes the QCD medium. The gluon transport coefficient $\hat{q}$ reflects the medium gluon density (say, its scattering property) while $L$ is the length of matter covered by the hard quark. Note that $\hat{q}$ can be very large in practice in a hot pion gas or quark gluon plasma~\cite{Baier:2002tc}. Energy loss of partons may thus be used to probe such dense media expected to be formed in high energy heavy ion reactions~\cite{Baier:1998kqBaier:1997kr,Gyulassy:1990yeGyulassy:1992xbWang:1992xyGyulassy:1994hrWang:1995fx}. The absolute values for $\omega_c$ in nuclei will be discussed below.

Using Mellin transform techniques, BDMS resumed the Poisson series to eventually give the quenching weight Eq.~(\ref{eq:poisson}) a simple integral representation.
This allows for the numerical computation of $D(\eps,\nu)$ from the BDMPS gluon spectrum~(\ref{eq:dIdo_out}) performed in~\cite{Arleo:2002kh}. The distribution has been given a simple analytic parameterization which we shall use in the present analysis.

\noindent \emph{Transport coefficient for nuclear matter $\hat{q}$}

The quenching of hadron spectra depends crucially on the transport coefficient for nuclear matter. The higher the $\hat{q}$, the larger the typical quark energy loss $\eps \propto \hat{q}$, hence the stronger the depletion in large nuclei. Based on the (not too) small $x$ gluon distribution in a nucleus, BDMPS provide a perturbative estimate for the transport coefficient in nuclear matter, $\hat{q}\simeq 0.25$~GeV/fm$^2$~\cite{Baier:1997sk}.

A slightly larger ---~although consistent~--- value has been obtained recently from our LO analysis of Drell-Yan production data~\cite{Arleo:2002ph} measured in $\pi^-$-A collisions by the NA3 collaboration,
\begin{equation}\label{eq:qhat}
\hat{q} = 0.72 \,\mathrm{GeV/fm}^2.
\end{equation}
This value will be taken for the numerical applications to come. It should already be kept in mind, however, that this value could not be determined with a high accuracy ($1 \sigma=0.54$~GeV/fm$^2$) because of the large statistical errors in the Drell-Yan measurements. 

As we shall discuss later (Section~\ref{se:discussion}), hadron depletion in DIS on light nuclei (say, Nitrogen and Neon targets) may provide an independent estimate for the transport coefficient $\hat{q}$. In particular, the high statistics HERMES data should prove of considerable help to constrain further the estimate~(\ref{eq:qhat}) from Drell-Yan production. 

\noindent \emph{Length $L$}

Once $\hat{q}$ is fixed, only the mean length $L$ covered by the hard quark in the nuclear medium remains to be determined. Assuming that hadronization occurs well after the hard quark has escaped from the nucleus, in other words when the hadron formation time $t_f$ is large, $L$ may simply be estimated for any given density profile $\rho({\bf r})$. Taking for simplicity a sharp sphere density $\rho({\bf r}) = \rho\times\Theta(R-|{\bf r}|)$, it is given by
\begin{equation}\label{eq:Lmax}
L_{\rm{max}} = \frac{3}{4}\, R,
\end{equation}
where the nuclear radius $R = (3/(4\pi\rho))^{1/3}\,A^{1/3}\simeq 1.17 A^{1/3}$~fm with $\rho = 0.15$~fm$^{-3}$.

When $t_f$ gets smaller than twice the nuclear radius, however, quarks produced on the ``back'' side of the nucleus hadronize while still inside the medium. The length~(\ref{eq:Lmax}) given above is therefore no longer correct. Integrating over the nuclear volume, we now obtain
\begin{equation}\label{eq:meanl}
L(t_f) \,=\, t_f\,\times\,\left[ 1 - \frac{3}{8}\,\frac{t_f}{R} + \frac{1}{64}\,\left(\frac{t_f}{R}\right)^3\right] \qquad \mathrm{if} \, \, t_f \leq 2\,R.
\end{equation}
Of course, $L(t_f) = L_{\rm{max}}$ at $t_f = 2\,R$. For small formation times (large nuclei), $t_f/R\ll 1$, energy loss effects are strongly reduced as the quark has propagated over shorter distances than (\ref{eq:Lmax}), $L(t_f \ll R) \simeq t_f \ll L_{\rm{max}}$. It is implicitly assumed here that a color singlet hadron ($C_A=0$ in (\ref{eq:dIdo_out})) does not lose any energy by gluon radiation. The produced hadron in the medium may nevertheless interact and be absorbed in the nucleus. In this approach ---~that we want to keep as simple as possible~---, this mechanism is not taken into account. However, we shall come back to this point in the discussion section.

The soft scales involved in the hadronization process prevent any perturbative calculation of the formation time $t_f$. Therefore, the absolute value for $t_f$ as well as its kinematic dependence is not clearly known. We shall take the Bialas-Gyulassy estimate~\cite{Bialas:1987cf} obtained in the framework of the Lund model\footnote{It has been stressed in~\cite{Accardi:2002tv} that Eq.~(\ref{eq:tf_lund}) slightly overestimates the Lund model predictions with standard parameters at small $z$. We checked however that this does not strongly affect our results.}
\begin{equation}\label{eq:tf_lund}
t_f = \left( \frac{\ln(1/z^2)-1+z^2}{1-z^2} \right) \times \frac{z\,\nu}{\sigma},
\end{equation}
where $\sigma$ is a non-perturbative scale in the order of the string tension, although slightly reduced in a nuclear environment~\cite{Accardi:2002tv}. We take $\sigma = 0.75$~GeV/fm throughout this study. The formation time (\ref{eq:tf_lund}) exhibits a maximum around $z \simeq 0.3$ and decreases like $t_f \sim (1-z)$ at large $z$, observed as well within gluon bremsstrahlung models~\cite{Kopeliovich:1995jt}.

\begin{figure}[h]
\begin{center}
\includegraphics[width=10.cm]{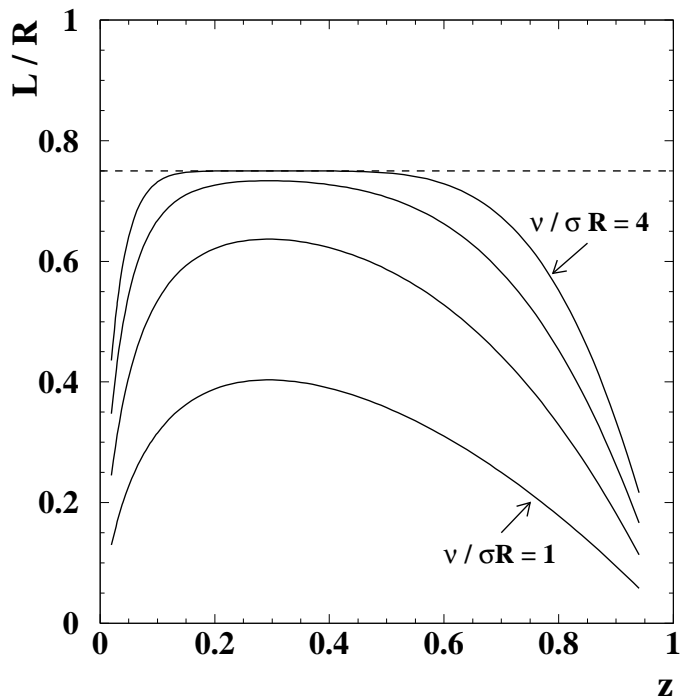}
\caption{Mean length (\ref{eq:meanl}) covered by the hard quark in the nucleus of radius $R$ as a function of $z$ and for different values of the quark energy $\nu$. The asymptotic length $L_{\mathrm{max}}$ assuming hadronization outside of the nucleus is shown as a dashed line.}
\label{fig:meanl}
\end{center}
\end{figure}

Plugging the Lund formation time (\ref{eq:tf_lund}) into Eq.~(\ref{eq:meanl}), the mean (normalized) length $L/R$ is plotted as a function of the quark energy $\nu$ and the hadron momentum fraction $z$ in Figure~\ref{fig:meanl}. Also shown as a dashed line is the asymptotic length $L_{\mathrm{max}}/R$ when hadronization occurs outside of the nucleus. We observe that $L$ gets much smaller than $L_{\mathrm{max}}$ when $z$ approaches 0 or 1, as $t_f$ vanishes. We shall thus expect in these specific kinematic limits a dramatic discrepancy with the prediction $L=L_{\mathrm{max}}$ assuming hadronization outside of the nucleus. Because of the Lorentz boost $L \propto \nu$ in (\ref{eq:tf_lund}), highly energetic quarks $\nu \gg \sigma R$ will fragment into hadrons after escaping the nuclear medium. As discussed in the next section and already apparent in Figure~\ref{fig:meanl}, finite formation time effects become irrelevant to describe high energy data.

To give the reader a feeling for such formation time effects, hadron attenuation will be computed in the following account assuming both the asymptotic length $L_{\mathrm{max}}$ (\ref{eq:Lmax}) (dashed lines) as well as the more realistic estimate Eq.~(\ref{eq:meanl}) (solid lines). 

\section{Results}\label{se:results}

Hadron production in semi-inclusive DIS on nuclear targets may now be determined from the perturbative expression (\ref{eq:multDIS}) using the nuclear fragmentation functions derived in the latter section. Calculations are  performed using the acceptance of the experiments. We have used the mean kinematic variables $\langle\nu\rangle$, $\langle z\rangle$, and $\langle Q^2\rangle$ measured experimentally whenever they were available, as it was shown in~\cite{Accardi:2002tv} that these were slightly underestimated (by less than 10\%) by the LO prediction (\ref{eq:multDIS}).

The quenching of hadron multiplicity $R_A^h(\nu)$ and $R_A^h(z)$ is compared to the EMC~\cite{Ashman:1991cx} and HERMES~\cite{Airapetian:2002ksMuccifora:2001znMuccifora:2002eaDiNezza} data on Cu as well as on N and Kr nuclei, respectively. We shall also compare in this section our results to the HERMES preliminary data on Neon~\cite{Elbakyan}. Finally, predictions on a Xenon target in the HERMES kinematic window are presented.

\subsection{$\nu$ dependence}

We start the section by first considering the quark energy dependence $\nu$ of the hadron quenching. As previously emphasized, the energy loss effects should vanish as the energy $\nu$ gets large as compared to the typical scale $\omega_c$ of the medium. Naturally, the quenching factor $R_A^h(\nu)$ should therefore increase with $\nu$ and eventually approaches unity at asymptotic energies. 

This is what is observed experimentally. In Figure~\ref{fig:nucu} is plotted the charged hadron attenuation $R^{h\pm}(\nu)$ in a Cu nucleus on a wide range, $15 \lesssim \nu \lesssim 200$~GeV. The calculations are found to be in good agreement with the measurements performed by the EMC collaboration. Notice that these high energy data do not exhibit any formation time effects in such a small nucleus, $\nu/\sigma R \gtrsim 6$, as shown from the tiny difference between the dashed and solid lines in Figure~\ref{fig:nucu}.

\begin{figure}[h]
\begin{center}
\includegraphics[width=9.2cm]{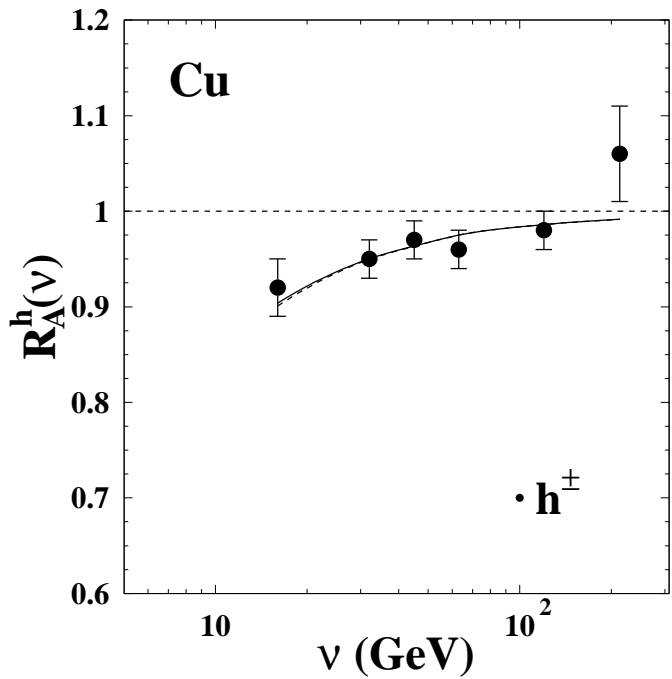}
\caption{Quenching of charged hadron spectra in Cu target as a function of $\nu$.  The theoretical calculations with ({\it solid}) and without ({\it dashed}) finite formation time effect are compared to EMC data~\cite{Ashman:1991cx}.}
\label{fig:nucu}
\end{center}
\end{figure}

Let us turn to lower energy data. The HERMES collaboration measured semi-inclusive hadron production on Nitrogen and Krypton targets. What is more, the experimental apparatus allows for a particle identification of various hadron species, such as pions, kaons, protons and anti-protons. Since no fragmentation functions are provided in the (anti-)proton channel, only the quenching of charged hadron, pion, and kaon spectra will be presented.

\begin{figure}[htbp]
\begin{center}
\includegraphics[width=14.cm]{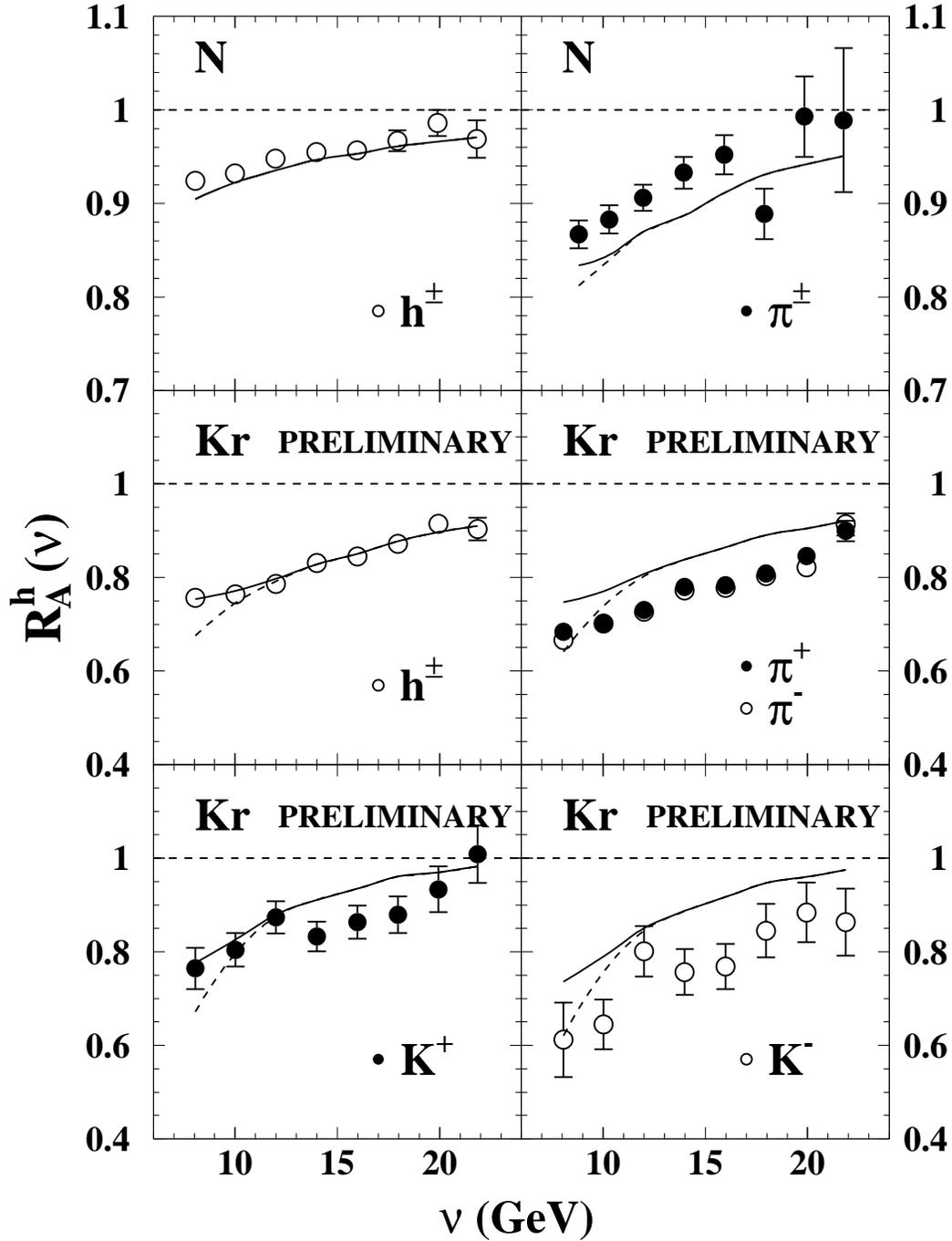}
\caption{Quenching of charged hadron, pion, and kaon spectra in N ({\it upper}) and Kr targets ({\it lower}) as a function of $\nu$. The theoretical calculations with ({\it solid}) and without ({\it dashed}) finite formation time effect are compared to HERMES data taken from Ref.~\cite{Airapetian:2002ksMuccifora:2001znMuccifora:2002eaDiNezza}. The 4\% systematic errors are not shown.}
\label{fig:nuntkr}
\end{center}
\end{figure}

Figure~\ref{fig:nuntkr} (upper panel) indicates that the quenching of charged hadrons on Nitrogen is extremely well reproduced by our predictions, while the calculations slightly overestimate the attenuation in the pion (isospin averaged) sector. It should be noted that the stronger pion quenching $R^\pi(\nu)<R^h(\nu)$ reflects only the different acceptance\footnote{More precisely, it comes from the different cuts in the $z$ acceptance. Pions were measured with $z>0.5$ while a cut $z>0.2$ has been applied to the charged hadron channel.} in the HERMES experiment. In the Kr target, the suppression of charged hadrons is again well described by the calculations. The trend predicted in the pion channel in Kr is close to what is observed experimentally, although the absolute magnitude of the pion quenching slightly underestimates the data ($\sim 5-8\%$).

Perhaps more interesting are the HERMES preliminary measurements on kaon quenching in krypton, $R_{\mathrm{Kr}}^{K\pm}(\nu)$ (figure~\ref{fig:nuntkr}, lower panel). A significantly larger depletion is observed in $K^-$ production on the whole energy range, $R^{K^-}(\nu) < R^{K^+}(\nu)$. Furthermore, this noticeable isospin dependence in the kaon sector is apparent in the theoretical predictions as well. It is indeed clear from Eqs.~(\ref{eq:suppDIS_approx}) to (\ref{eq:modelFF}) that the quenching is directly related to the slope of the vacuum fragmentation functions\footnote{We omit, for clarity, the integral $\int d\eps D(\eps)$ in Eq.~(\ref{eq:suppDIS_approx2}).}
\begin{eqnarray}\label{eq:suppDIS_approx2}
R_A^{h}(\nu) & \approx & 1+ \frac{\partial D_u^h}{\partial z} \frac{1}{D_u^h(z)} \frac{z\,\eps}{\nu} \nonumber \\
& \approx& 1 - \eta_u^h \times \frac{z\,\eps}{\nu (1-z)}
\end{eqnarray}
where $\eta_u^h$ is the slope of the fragmentation functions at large $z$, $D_u^h(z, Q^2) \sim (1-z)^{\eta_u^h}$. Schematically, the up quark will preferably fragment into a $K^+$ by picking up easily a surrounding $\bar{s}$ quark in the sea (valence-type fragmentation channel). Inversely, the difficulty to produce a negative kaon from a hard up quark (sea-type) will translate into a steeper slope, $\eta_u^{K^-}=\eta_u^{K^+}+1$ and thus in a stronger quenching Eq.~(\ref{eq:suppDIS_approx2}). At very small $x \ll 1$, semi-inclusive production originates from the sea quark fragmentation, $u=\bar{u}$~\footnote{Furthermore, the sea is almost flavor symmetric $\bar{u}\simeq\bar{d}$ at small $x$. However, down quarks will be disfavored due to their electric charge.}. We therefore predict that no more isospin effect should be observed in this region, and thus $R^{K^-}(\nu)\simeq R^{K^+}(\nu)$

\begin{figure}[htbp]
\begin{center}
\includegraphics[width=14.cm]{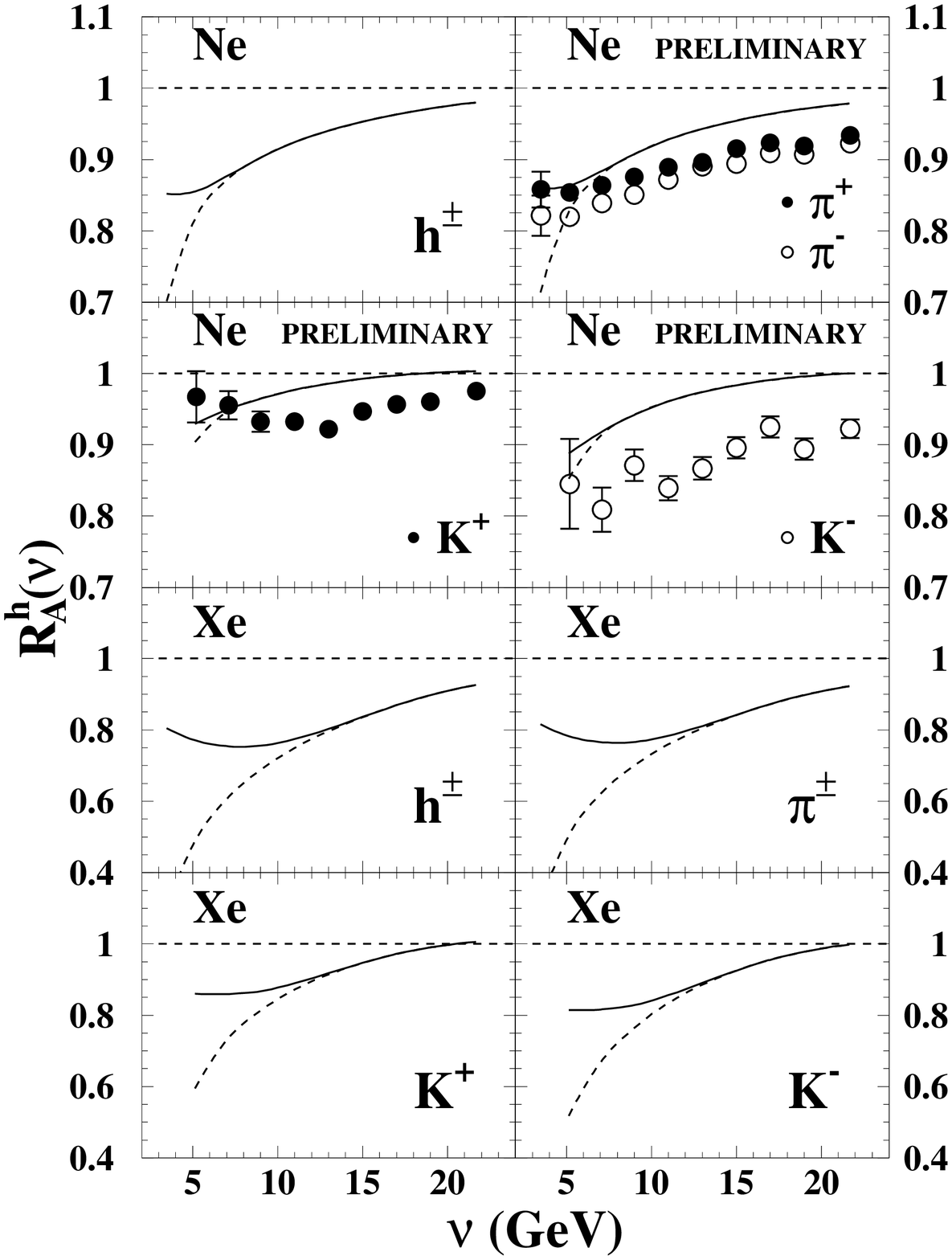}
\caption{Quenching of charged hadron, pion, and kaon spectra in Ne ({\it upper}) and Xe targets ({\it lower}), with ({\it solid}) and without ({\it dashed}) finite formation time effect, as a function of $\nu$. The HERMES preliminary data on Ne nuclei are taken from Ref.~\cite{Elbakyan}. The 3.3\% systematic errors are not shown.}
\label{fig:nunexe}
\end{center}
\end{figure}

Formation time effects start to become visible in Krypton targets when $\nu$ is not too large, $\nu\lesssim 12$~GeV (see Figure~\ref{fig:nuntkr}). Nevertheless, the rather small differences between the dashed and solid curves indicate that the krypton nucleus (A=84) is not too large to probe the hadronization {\it in} the medium. More pronounced effects are expected in heavier nuclei.

Finally, hadron attenuation has been computed on a Neon target (Figure~\ref{fig:nunexe}, {\it upper panel}) measured very recently by the HERMES collaboration~\cite{Elbakyan}. While the observed pion quenching is well reproduced by the calculations, the predicted attenuation noticeably overestimates the preliminary data in the kaon sector, and particularly in the $K^-$ channel. This feature was already present, although less pronounced, in the comparison with Krypton measurements. The underestimate of the slopes of the flavor-tagged $u \to K^+ (K^-)$ fragmentation functions, poorly constrained by $e^+e^-$ data, may be at the origin of this discrepancy~\cite{Kretzer:2000yf}. In particular, the assumption made in Ref.~\cite{Kretzer:2000yf} that the sea type channels (e.g. $D_u^{K^-}$) get suppressed by one extra power in $(1-z)$ (see discussion above) seems to be disfavored by experimental data on kaon production~\cite{Arneodo:1989ic}. 

Figure~\ref{fig:nunexe} (lower panel) also displays our predictions on Xenon nuclei (A=131), in which formation time effects get larger. As later stressed in the discussion section, measurements in such a heavy nucleus will be of crucial importance to disentangle the possible various mechanisms responsible for the observed hadron attenuation.

\subsection{$z$ dependence}

Another way to probe the dynamics of hadronization in a nuclear medium is to look at the $z$ dependence of hadron quenching. In our calculations, we expect a stronger hadron quenching as $z$ increases for two reasons. First, as can been seen from the rough estimate Eq.~(\ref{eq:suppDIS_approx2}), the slope of the fragmentation functions increases with $z$, and thus so does the attenuation $R_A^h(z)$. 

The second reason that leads to significant hadron depletion at large $z$ actually comes from the restricted phase space. Measuring hadrons with large momentum fraction $z$ selects on those (rare) fragmenting quarks that have lost a tiny amount of energy. Assuming for simplicity that the fragmentation functions are flat, $z^* D(z^*,Q^2)\simeq z D(z,Q^2)$, the quenching factor may be approximated by
\begin{equation}\label{eq:ps}
R(\nu,z) \simeq \int_0^{(1-z)\nu} d\eps \,D(\eps,\nu),
\end{equation}
which vanishes as $z$ approaches 1. Let us note however that the ratio (\ref{eq:ps}) will tend to a constant $p_0>0$ in the parton energy loss frameworks developed by Gyulassy, L\'evai, Vitev~\cite{Gyulassy:1999zdGyulassy:2000fs} and Wiedemann~\cite{Wiedemann:2000zaWiedemann:2000tf} in which the quenching weight $D(\eps,\nu)$ contains a discrete part, $p_0 \delta(\eps)$~\cite{Salgado:2002cdSalgado:2003gb}. Similar behavior is also observed within the BDMPS formalism when ultra soft gluons are removed from the perturbative spectrum~\cite{Arleo:2002kh}.

\begin{figure}[h]
\begin{center}
\includegraphics[width=9.2cm]{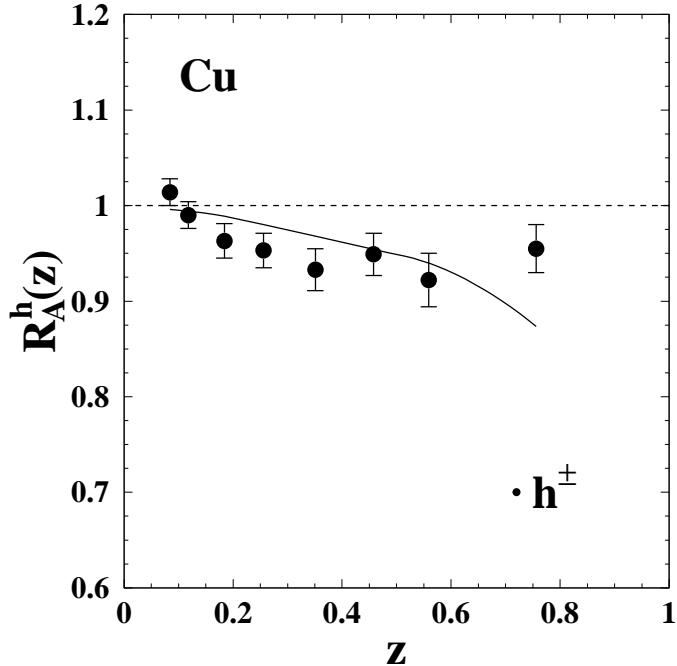}
\caption{Quenching of charged hadron spectra in Cu target as a function of $z$.  The theoretical calculations with ({\it solid}) and without ({\it dashed}) finite formation time effect are compared to EMC data~\cite{Ashman:1991cx}. The dashed and solid lines are superimposed.}
\label{fig:zcu}
\end{center}
\end{figure}

\begin{figure}[htbp]
\begin{center}
\includegraphics[width=14.cm]{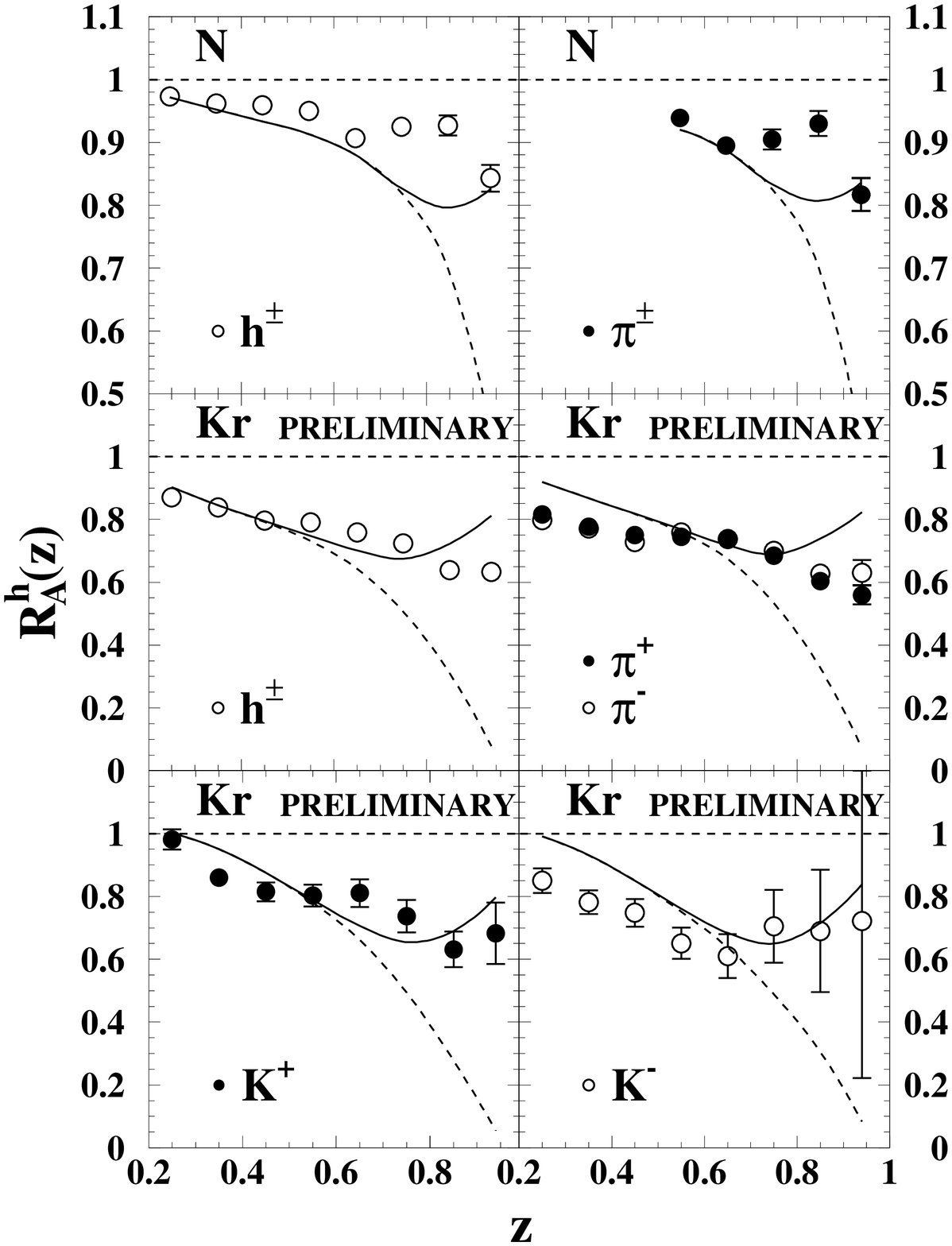}
\caption{Quenching of charged hadron, pion, and kaon spectra in N ({\it upper}) and Kr targets ({\it lower}) as a function of $z$. The theoretical calculations with ({\it solid}) and without ({\it dashed}) finite formation time effect are compared to HERMES data taken from Ref.~\cite{Airapetian:2002ksMuccifora:2001znMuccifora:2002eaDiNezza}. The 4\% systematic errors are not shown.}
\label{fig:zntkr}
\end{center}
\end{figure}

Looking at Figure~\ref{fig:zcu}, our predictions on the $z$ dependence of the charged hadron quenching $R^{h^\pm}(z)$ in Copper targets describe well the EMC measurements, with the exception of the highest $z$ bin. Again, high energy quarks hadronize well after escaping the nucleus in this whole $z$ range.

On the contrary, we do observe huge formation time effects in the HERMES data which are taken at much lower energy $\langle\nu\rangle \simeq 10$--$20$~GeV (Figure~\ref{fig:zntkr}). The predictions fairly reproduce the HERMES measurements for all hadron species when $z$ is not too high. At large $z\gtrsim 0.7$, nevertheless, the discrepancy between the predictions assuming $L=L_{\mathrm{max}}$ (dashed lines) and the data is striking. As discussed above, the hadron attenuation $R(z)$ falls too steeply as $z$ goes to one, unlike the experimental results. This does not come as a surprise as we have already anticipated in the latter section that this assumption is no longer justified since $L \simeq t_f \sim (1-z)\nu/\sigma \ll L_{\mathrm{max}}$.

When the realistic length Eq.~(\ref{eq:meanl}) is used, the typical scale $\omega_c \propto L^2 \sim (1-z)^2$ is getting smaller and weakens the quark energy loss ---~and subsequently the hadron attenuation~--- as $z$ increases. This effect somehow compensates the quenching due to the slope of the fragmentation functions (\ref{eq:suppDIS_approx2}) and to phase space restriction (\ref{eq:ps}) at moderate $z$, and eventually takes over at $z\gtrsim 0.9$ (see Figure~\ref{fig:zntkr}). This non-monotonic behavior of the hadron depletion is actually a manifestation of the quadratic behavior of the parton energy loss in QCD media. We may indeed convince ourselves of this by replacing the energy loss $\eps\propto \omega_c \sim (1-z)^2$ in the rough estimate (\ref{eq:suppDIS_approx2}). It is however doubtful whether this could seen in the data as the final state interaction of the produced hadron may somehow modify this picture. As shown in Figure~\ref{fig:zntkr}, a nice agreement is reached between these calculations (solid lines) and the HERMES data for all $z$. 

The predicted hadron attenuation in Ne and Xe targets is also plotted in Figure~\ref{fig:znexe} as a function of $z$. The preliminary HERMES measurements in Neon are well described by these predictions. Once more, let us emphasize that measurements in Xenon targets will prove essential to further constrain the hadron formation time estimates at large~$z$.

\begin{figure}[htbp]
\begin{center}
\includegraphics[width=14.cm]{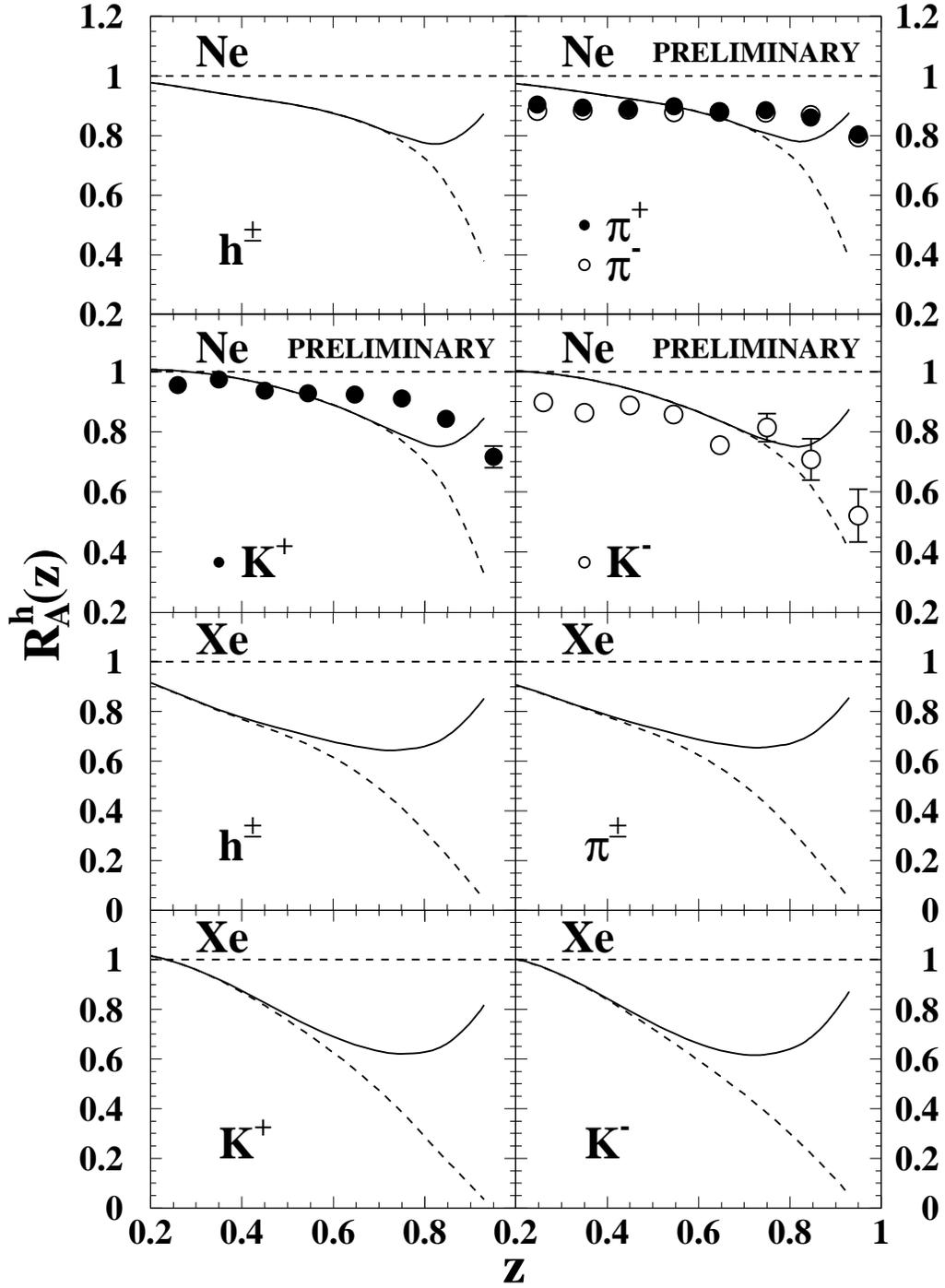}
\caption{Quenching of charged hadron, pion, and kaon spectra in Ne ({\it upper}) and Xe targets ({\it lower}), with ({\it solid}) and without ({\it dashed}) finite formation time effect, as a function of $z$. The HERMES preliminary data on Ne nuclei are taken from Ref.~\cite{Elbakyan}. The 3.3\% systematic errors are not shown.}
\label{fig:znexe}
\end{center}
\end{figure}

Let us finally stress that theoretical uncertainties might get large in the high $z\to 1$ region. In particular, important higher twist corrections may come into play in this restricted region of phase space~\cite{Wang:2002ri,Accardi:2002tv}, given moreover the fact that the typical values $Q^2 \simeq 2 - 4$~GeV$^2$ measured in these reactions are pretty small. Moreover, the resummation of soft and collinear gluons leads to large contributions $\alpha_s \log(1-z)/(1-z)$ in the fragmentation functions~\cite{Cacciari:2001cw} which are not taken into account in the leading order analysis~\cite{Kretzer:2000yf} we use here.

\section{Discussion and summary}\label{se:discussion}

The nuclear dependence of semi-inclusive hadron production in DIS comes in our calculations from the sole effect of the quark energy loss in nuclei. Of course, many other mechanisms, which have not been considered here, could play a role in the hadron quenching observed experimentally. The absorption of the (pre)hadrons produced in the medium may for instance substantially affect its production~\cite{Bialas:1983kn,Bialas:1987cf,Accardi:2002tv,Falter:2003di}. This will be particularly true at large $z$ and/or small $\nu$ when the formation time is small. It was also stressed that the partial deconfinement which could occur in large nuclei rescales the fragmentation functions and eventually suppress the hadron yields~\cite{Jaffe:1984zwClose:1985znNachtmann:1984py,Accardi:2002tv}. 

However, although such effects have not been taken into account here, a very nice description of the EMC and more recent HERMES preliminary data for several hadron species and various nuclei is still achieved. The already good agreement assuming only energy loss effects in our approach is therefore a hint that the above mentioned competing processes play a rather modest role. Let us moderate this statement. The strength of each of these mechanisms is governed by one parameter which is in general poorly constrained. Take the quark energy loss for instance. We have used here the transport coefficient for nuclear matter $\hat{q}=0.72$~GeV/fm$^2$ extracted from Drell-Yan data, whose absolute magnitude is not known with a great accuracy. Similarly, the nuclear absorption strongly depends on the pre-hadron cross section $\sigma^*$ about which little is known.

It is therefore a delicate matter to disentangle these various processes. We suggest a couple of observables which may help us in this prospect. First, we have seen that the mean length $L \simeq t_f$, hence the scale $\omega_c$, becomes independent of $R$ when the nuclear radius is large, $R \gg \nu/\sigma$.  The effects of quark energy loss on hadron quenching thus somehow saturates when the nuclei are sufficiently heavy. This can already be seen from the comparison of the rather similar hadron quenching in Kr and Xe targets, $R_{\mathrm{Kr}} \simeq R_{\mathrm{Xe}}$ (Figures~\ref{fig:zntkr} and \ref{fig:znexe}, solid lines). It is clearly at variance with the absorption models which predict of course a much stronger attenuation as the nucleus gets larger. We therefore predict that the ratio of hadron depletion in two large nuclei (say, Xe/Kr) will provide a good measurement of hadron absorption, {\it only}. Inversely, hadron quenching in small nuclei could help to constrain the transport coefficient in nuclear matter, without being spoiled by hadronic absorption\footnote{Let us however mention that rescaling models predict a non-negligible suppression in such small nuclei~\cite{Accardi:2002tv}.}. Fitting $\hat{q}$ to the charged hadron and pion quenching in Nitrogen and Neon data, we found $\hat{q}=0.59\pm 0.18$~GeV/fm$^2$ and $\hat{q}=0.70 \pm 0.06$~GeV/fm$^2$, respectively, in excellent agreement with the Drell-Yan estimate\footnote{This consistency between Drell-Yan and DIS data was discussed in Ref.~\cite{Arleo:2002ki}}, Eq.~(\ref{eq:qhat}), which suffered from a larger statistical uncertainty.  

We also think that the isospin dependence of the kaon quenching could be important. As already mentioned, a similar attenuation ratio $R^{K^-}(\nu,z)\simeq R^{K^+}(\nu,z)$ is expected at small $x$ if quark energy proves the most relevant mechanism at work. On the contrary, a stronger $K^-$ quenching should still be observed in the framework of absorption models from the larger interaction of negative kaons. The $Q^2$ dependence (at a given $\nu$) of the ratio $R^{K^+}/R^{K^-}$ may thus clarify the relevant underlying processes.

To conclude, a leading order calculation of semi-inclusive hadron production in DIS on nuclei has been performed. The fragmentation functions were modified to take into account the energy loss incurred by hard quarks while they propagate through the nuclear medium. The quenching of hadron spectra was then compared to the EMC and HERMES preliminary data in several nuclear targets, as a function $\nu$ and $z$. Using the transport coefficient determined from Drell-Yan production, a nice agreement between our predictions and the experimental measurements is observed for the various hadron species. In particular, we emphasized that hadron production in small nuclei may help us to determine and constrain what the transport coefficient (hence the quark energy loss) of nuclear matter is. Measurements of semi-inclusive hadron production in heavy nuclei are highly desirable since, in addition to be a sensitive probe of formation time effects, these prove crucial to disentangle quark energy loss on the one hand from hadron absorption on the other hand. 

Finally, it should be interesting to constrain, from the HERMES and EMC data, the rapidity dependence of the quenching of charged hadron and pion hadroproduction inclusive spectra in deuteron-gold collisions at RHIC energy~\cite{Adler:2003iiBack:2003nsAdams:2003im}.

\subsection*{Acknowledgments}

I would like to thank N\'estor Armesto, Valeria Muccifora and Jianwei Qiu for discussions and many useful comments about this study, and Garegin Elbakyan for providing me with the HERMES preliminary Neon data.

\end{document}